\documentclass{llncs}

\newif\iftr
\trfalse 

\usepackage{paralist}
\usepackage{float}
\usepackage{wrapfig}
\usepackage[labelformat=simple]{subcaption}
\usepackage{tikz}
\usepackage{booktabs}
\usepackage[override]{cmtt}
\usepackage{algorithm}
\usepackage[inline]{enumitem}
\usepackage{algpseudocode}
\usepackage{algorithmicx}
\usepackage{boxedminipage}
\usepackage{multirow,array,varwidth}
\usepackage{url}
\usepackage{pdfpages}
\usepackage{graphicx}
\usepackage{color}
\usepackage[sort]{cite}
\usepackage[bookmarks=false]{hyperref}
\usepackage{xspace}
\usepackage{listings}
\usepackage{amsmath,amstext,amssymb,amsfonts,latexsym}

\usepackage{titlesec}
\titlespacing{\section}{0pt}{*1.7}{*1.3}
\titlespacing{\subsection}{0pt}{*1.7}{*0.8}

\renewcommand{\paragraph}[1]{\vspace{3pt}\noindent\textbf{#1}}

\algtext*{EndFor}
\algtext*{EndIf}
\algtext*{EndWhile}

\newcommand{\coin}{Aspen}
\newcommand{\SOS}{Service-Oriented Sharding}
\newcommand{\sos}{service-oriented sharding}
\newcommand{\Sos}{Service-oriented sharding}
\newcommand{\ssn}{service number}

\newcommand{\ssns}{service numbers}
\newcommand{\pch}{payment channel}
\newcommand{\pchs}{payment channels}
\newcommand{\chp}{service protocol}
\newcommand{\chps}{service protocols}

\newcommand{\fp}{funding pore}
\newcommand{\fps}{funding pores}

\newcommand{\bud}{bud}
\newcommand{\buds}{buds}

\newcommand{\cc}{registration channel}

\graphicspath{ {./figures/} }

\begin{document}

\title{
\SOS\ with \coin
\vspace{-5mm}
}

\author{
{
Adem Efe Gencer\qquad Robbert van Renesse\qquad Emin G\"un Sirer\\
{\small 
Initiative for CryptoCurrencies and Contracts (IC3)\\[-1pt]
Computer Science Department, Cornell University
}}
}

\institute{\vspace{-8mm}}
\maketitle

\begin{abstract} 

The rise of blockchain-based cryptocurrencies has led to an explosion of 
services using distributed ledgers as their underlying infra-structure. 
However, due to inherently single-service oriented blockchain protocols, 
such services can bloat the existing ledgers, fail to provide 
sufficient security, or completely forego the property of trustless 
auditability. Security concerns, trust restrictions, and scalability limits 
regarding the resource requirements of users hamper the sustainable 
development of loosely-coupled services on blockchains.

This paper introduces \coin, a sharded blockchain protocol designed to 
securely scale with increasing number of services.
\coin\ shares the same trust model as Bitcoin in a peer-to-peer network 
that is prone to extreme churn containing Byzantine participants.
It enables introduction of new services without compromising the 
security, leveraging the trust assumptions, or flooding users with 
irrelevant messages.

\end{abstract} 

\section{Introduction}\label{introduction}

Blockchains offer many opportunities for facilitating innovation
in traditional industries. They have received extensive attention 
due to the trustless auditability, tamper-resistance, and transparency 
they provide in networks with Byzantine participants.
Not surprisingly, there is much commercial interest in
developing specialized blockchain solutions~\cite{coindesk2016state}. 
There have been proposals to use blockchains as an underlying layer 
for services including 
managing digital assets~\cite{coloredCoins2016}, issuing 
securities~\cite{chain-com}, maintaining land records and 
deeds~\cite{georgian-land-records-on-the-blockchain, benben2016}, tracking 
intellectual property~\cite{omi2016, ascribe2016, uproov2016}, facilitating 
online voting~\cite{voting}, registering domain names~\cite{namecoin}, as 
well as others. Ongoing projects explore ways of making it easier to build 
such services through Blockchain-as-a-Service~(BaaS)
platforms~\cite{baas-ms, baas-ibm}.

This movement, towards increased adoption of blockchains for specialized
purposes, portends a dangerous trend: accommodating all of these diverse
uses, either in a single blockchain or in separate blockchains, inherently 
requires complex tradeoffs. The simplest approach, of layering these
additional blockchains on top of an existing, secure blockchain with
sufficient mining power to withstand attacks, such as 
Bitcoin~\cite{nakamoto2008bitcoin}, leads to a stream of costly and
burdensome transactions. Indeed, we have seen the controversial 
\texttt{OP\_RETURN}
opcode adopted for this purpose, and its use has been increasing 
rapidly~\cite{opreturn_stats},
in line with increased usage of layered blockchains. Yet these transactions,
which use Bitcoin solely as a timestamping and ordering service, 
increase the resource requirements for system participation and the time to
bootstrap a node. In contrast, creating a dedicated, specialized, 
standalone blockchain avoids this problem, but suffers from 
a lack of independent mining power to secure the infrastructure.
Duplicating the mining infrastructure used to secure Bitcoin is not only
costly and environmentally unfriendly, but it is difficult to bootstrap
such a system. Faced with this dilemma, some have turned to permissioned 
ledgers with closed participants~\cite{hyperledger2016, corda}, forgoing the 
open architecture, the flexible trust model, and the strong security guarantees 
of the existing Bitcoin mining ecosystem.

In this paper, we present \coin, a protocol that securely shards the blockchain 
to provide high scalability to service users. This protocol employs a sharding 
approach that comes with the following benefits:
\begin{enumerate*}[label=(\arabic*)]
\item preserves the total computational power of miners to secure the whole 
blockchain,
\item prevents users from double-spending their funds while maintaining the 
same trustless setup assumptions as Bitcoin,
\item improves scalability by absolving non-miner participants -- i.e. service 
users -- from the responsibility of storing, processing, and propagating 
irrelevant data to confirm the validity of services they are interested in. 
\end{enumerate*}
In this protocol, a coffee shop does not have to worry about the land and 
deed records in the blockchain to validate the payment system.

Sharding is a well-established technique to improve scalability by 
distributing contents of a database across nodes in a network.
But sharding blockchains is non-trivial. The main difficulty is to preserve 
the trustless nature while hiding parts of a blockchain from other 
nodes. It is an open research question whether it is possible to 
shard blockchains in a way that the output of a transaction in one shard 
can be spent at another while still satisfying the trustless validation of 
transaction history.
In this work, the key insight behind sharding the blockchain is to distribute 
transactions to blocks with respect to services they are used for.

This paper outlines \emph{\sos}, a technique for sharding block-chains 
that promises higher scalability and extensibility without modifying 
Bitcoin's trust model. It instantiates this technique in \emph{\coin}, a 
blockchain sharding protocol that expedites user access to relevant services,
makes service integration and maintenance easier, and achieves better fairness 
while demanding only a fraction of resources from users.

\section{\SOS}\label{\sos}

The core idea behind \sos\ is to partition a blockchain such that users can 
fully validate the correct functioning of their services 
\begin{enumerate*}[label=(\arabic*)]
\item without relying on trusted entities and
\item while keeping track of only the subset of the blockchain that is relevant 
to their services.
\end{enumerate*}
This technique shares the same network and trust model as Bitcoin and related 
cryptocurrencies. 
\Sos\ is built around a multiblockchain structure, where multiple chains are rooted in the same genesis block and share common checkpoints as shown in Fig.~\ref{fig:drop}.
Building blocks that comprise \sos\ can be summarized as follows:

\begin{figure}[t]
\centering
\includegraphics[width=\linewidth]{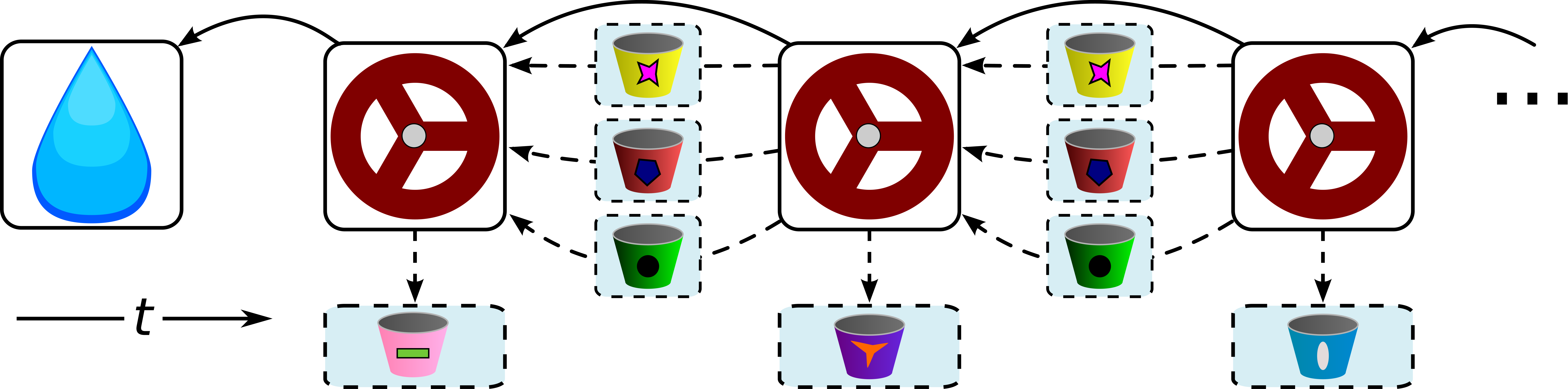}
\vspace{-1.2\baselineskip}
\caption{Multiblockchain structure of \sos. Each channel contains the same 
genesis block (drop) and checkpoints (valves), as well as the exclusive 
transactions of a specific service (buckets with the same symbol). Generating a 
checkpoint requires a proof of work. Miners distribute transactions to 
designated blocks (a subset of dashed rectangles) secured by 
checkpoints.}\label{fig:drop}
\end{figure}

%
%
%

\paragraph{Channel.} A chain in a blockchain built on a shared genesis block 
containing
\begin{enumerate*}[label=(\arabic*)]
\item all transactions of a specific service, and
\item common checkpoints involving transactions for the overall management of services.
\end{enumerate*}
For instance, a domain name resolution service would use a dedicated channel to 
store custom transactions in the form of DNS resource records. Such 
transactions are kept separate from common checkpoints. Hence, services are 
loosely coupled and resilient to changes.

\Sos\ handles requests associated with a certain service by annotating each 
channel and the corresponding transactions with the same unique identifier, 
called \emph{\ssn}. Two special channels, \emph{payment} and 
\emph{registration}, are defined by the system and help bootstrap the network. 
The default service that enables users to exchange funds runs on the \pch, and 
the \cc\ is used to add or update services.
Users store, process, and propagate transactions on channels only for the 
relevant services.

\paragraph{Protocol.} A set of rules regarding services and their integration. 
A \emph{\chp} defines the validity of transactions in a given channel. It describes:
\begin{enumerate*}[label=(\arabic*)]
\item the syntax for each transaction type,
\item the relationship between transactions within a channel,
\item the size, frequency, and format constraints for blocks that keep 
transactions.
\end{enumerate*}
The \emph{integration protocol} specifies the security, incentive mechanism, valid \ssns, the genesis block, and the inter-channel communication process between the \pch\ and the other channels.

%

\paragraph{Transaction.} The smallest unit of data for adding content to a
channel. Transactions are grouped into blocks and appended to each channel according to their \ssn. A block is valid if it
\begin{enumerate*}[label=(\arabic*)]
\item embodies valid transactions sharing the same \ssn\ and
\item complies with the integration protocol and the relevant \chp.
\end{enumerate*}

\paragraph{Service Integration and Maintenance.} The process of introducing 
services and updating the existing ones. \Sos\ resolves this process completely 
on the blockchain in three phases. First, users propose protocols to introduce 
or update services by generating transactions for the \cc. 
Each such transaction contains a set of \chps\ with distinct \ssns. A \chp\ is 
specified in a platform independent language such as 
WebAssembly~\cite{webAssembly} or Lua~\cite{lua}. In the second phase, miners 
conduct an election to choose a \cc\ transaction. This transaction specifies the protocols that miners are collectively willing to adopt. Miners indicate 
their choice using \emph{ballots}. A ballot is a transaction that 
contains a reference to a particular transaction in the \cc. Each ballot is 
part of a checkpoint that requires a proof of work to generate. This provides
\begin{enumerate*}[label=(\arabic*)]
\item representation proportional to mining power, and
\item protection against censorship of ballots.
\end{enumerate*}
Finally, if a particular transaction is referred by more than a certain 
fraction $\tau$ of ballots, its protocols become active. An active \chp\ determines the validity of new transactions added to the corresponding channel.

This process enables evolutionary refinement with the confidence of 
sustainability. Users are involved in the process through their 
proposals. The election mechanism ensures that the majority of the mining power 
intends to serialize transactions for the new or updated services.

\section{\coin}\label{\coin}

\begin{figure}[t]
\centering
\includegraphics[width=\linewidth]{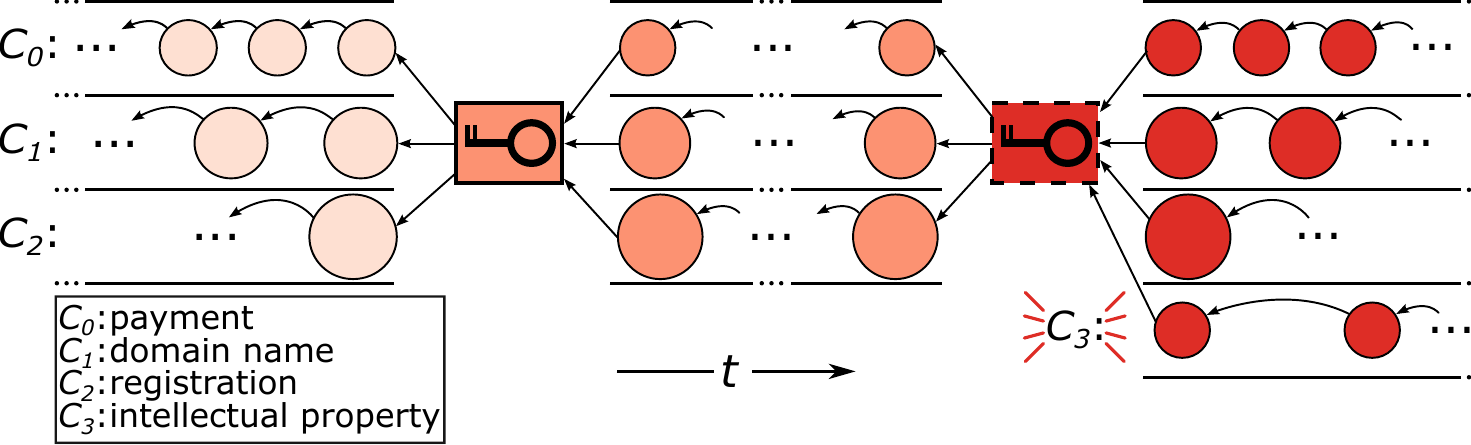}
\vspace{-1.2\baselineskip}
\caption{Structure of the \coin\ chain. Upon generating a key block shared by 
all channels, a miner serializes service-specific transactions only in the 
corresponding microblock (circles) chains. Shading indicates blocks generated 
by a specific miner. A \bud\ (dashed key block) introduces the intellectual 
property service.}\label{fig:chain}
\end{figure}

While \sos\ can be built on any blockchain protocol~\cite{nakamoto2008bitcoin, 
bitcoinNg, byzcoin, ethereum}, we instantiate on Bitcoin-NG~\cite{bitcoinNg}, a blockchain protocol that improves 
transaction throughput and consensus latency of Bitcoin under the same trust 
model. The protocol makes the following changes with \sos:


\paragraph{Multiple Microblock Chains.} Traditional blockchain protocols 
strive to agree on a single \emph{main chain} in which all transactions are 
totally ordered. However, not all transactions are related or even need such an 
ordering. This leads to a seemingly irreconcilable tradeoff between the 
scalability of independent blockchains and the security of monolithic ones. 
The central idea behind \coin\ is to resolve this conundrum by having a series 
of independent microblock chains conjoined at common key blocks. A channel represents 
the combination of the same genesis block, all key blocks, and the set of 
microblock chains containing custom transactions annotated with the same \ssn. 
Fig.~\ref{fig:chain} illustrates the structure.

Each channel maintains key blocks to enforce the integration protocol. To 
prevent bloating key blocks, \coin\
\begin{enumerate*}[label=(\arabic*)]
\item limits the number of channel references in a key block and
\item omits references to non-\pchs\ with no transactions on their 
latest microblock chain -- i.e. \emph{inert channels}.
\end{enumerate*}
Note that users can fully validate an inert channel service using key blocks of the \pch.



\paragraph{Extensibility.} \coin\ updates or introduces services at 
designated growth points, called \emph{\buds} (See Fig.~\ref{fig:chain}). A 
\bud\ is a key block at a protocol-defined height in terms of the number of key 
blocks from the genesis block. \coin\ adopts proposals based on ballots in key blocks between the current and the preceding \bud.


\begin{figure}[t]
\centering
    \begin{subfigure}[t]{.48\textwidth}
        \centering
        \includegraphics[height=2.5cm]{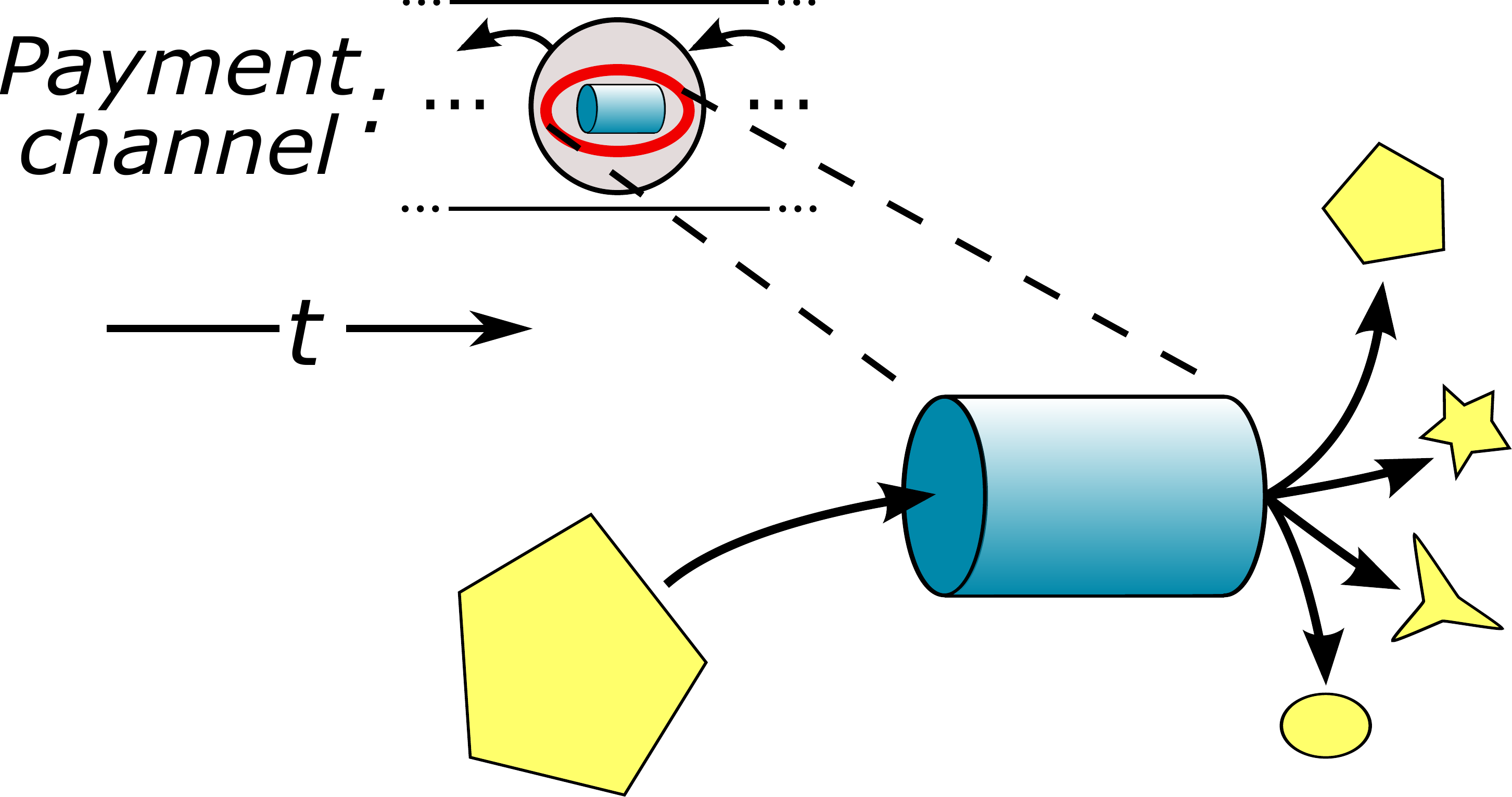}
\vspace{-0.1\baselineskip}
        \caption{A \fp.}\label{fig:pore}
    \end{subfigure}
    ~
    \begin{subfigure}[t]{.48\textwidth}
        \centering
        \includegraphics[height=2.5cm]{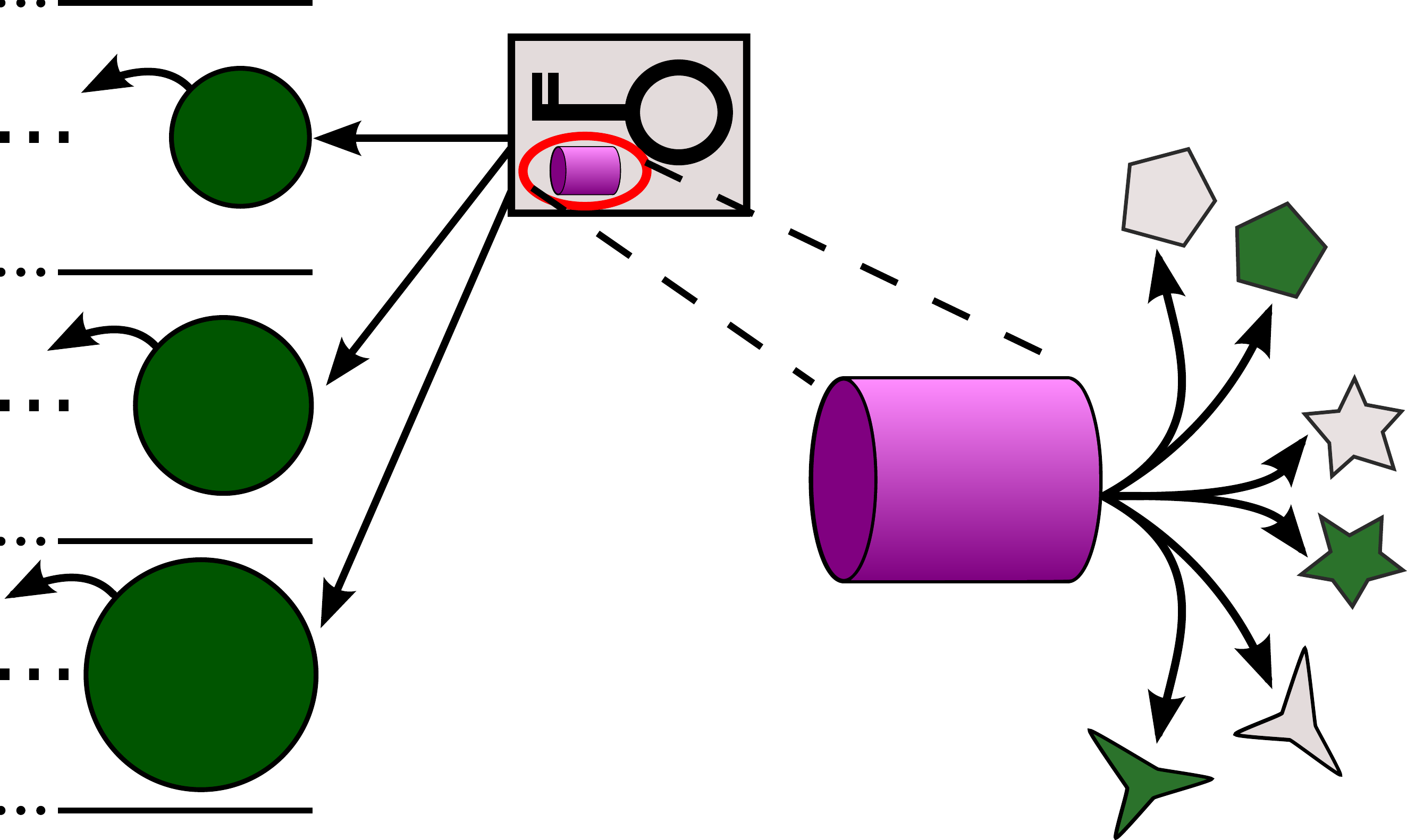}
\vspace{-0.1\baselineskip}
        \caption{A coinbase transaction.}\label{fig:coinbase}
    \end{subfigure}
\vspace{-0.4\baselineskip}
\caption{
\subref{fig:pore}~A \fp\ (cylinder) makes \pch\ outputs (pentagons) spendable at specific channels.
\subref{fig:coinbase}~Rewards are split between the current and the previous miner for each channel.
}\label{fig:transactions}
\end{figure}

\paragraph{Flow of Funds} \coin\ enables users to detect double spends by 
making each fund spendable only in a specific channel. A special \pch\ 
transaction, \emph{\fp}, enables users to lock funds to other channels. A \fp\ 
annotates each output with the \ssn\ of an existing destination channel where 
it can be spent. Note that transfers across channels are allowed only in one 
way, from the \pch\ to others. Fig.~\ref{fig:pore} illustrates a \fp.

Alternatively, users can directly buy locked funds at the target channel to pay 
for the service running on the corresponding channel. The protocol enforces a 
high minimum fee for serializing \fps\ to
\begin{enumerate*}[label=(\arabic*)]
\item discourage users from bloating the \pch\ and 
\item improve the fungibility of funds in non-\pchs.
\end{enumerate*}
Contrary to Bitcoin's \texttt{OP\_RETURN} transactions, this process does not 
yield any unspendable outputs.

Following sections detail the incentive and security mechanisms in \coin.

\subsection{Reward Structure}\label{reward-structure}

The process of keeping the complete blockchain, serializing transactions, and 
securing the system consumes miner resources. \coin\ uses a Bitcoin-like 
cryptocurrency to encourage miners to continue facilitating this costly 
process. A coinbase transaction in a key block provides separate outputs to 
compensate the current and the previous miner for each service they provision. 
Each output indicates the source channel of rewards where funds can be spent 
(See Fig.~\ref{fig:coinbase}).

\paragraph{Generating Key Blocks.} Miners receive a fixed subsidy for each key block they generate as an incentive for using their mining power to secure the blockchain and facilitating the voting process of service integration and maintenance.

\paragraph{Serializing Transactions.} Each \chp\ specifies the validity 
requirements for its transactions. The common property of all transactions is a 
fee that miners collect for adding them to the corresponding microblocks.

\paragraph{Extending the Longest Chains.} As an incentive for the next miner to attach her key block to the latest microblock~\cite{bitcoinNg}, \coin\ distributes fees between the current miner and the next one for each microblock chain.

\paragraph{Extending Multiple Chains.} Miners can spend transaction fees only 
in the corresponding channels that they were collected from. The high minimum 
fees for \fps\ encourage users to purchase locked funds. Hence miners gain 
additional incentives to serialize non-\pch\ transactions.

\subsection{Security}\label{security}

The following properties are critical to the security of a blockchain 
protocol.

\paragraph{Authenticity.} The property of having an indisputable origin.
Transactions require a set of cryptographic signatures to prove the ownership of funds that are used as fees. Hence, provided that it is infeasible to forge signatures, pseudonymous identities cannot deny committing transactions.

\paragraph{Irreversibility.} The protection against overwriting or deleting
transactions. Double spending is an instance of violating this property. Malicious miners may modify or remove a set of transactions from a blockchain by updating some common prefix with 
different blocks -- i.e. forks.
\coin\ secures the blockchain against
\begin{enumerate*}[label=(\arabic*)]
\item key block forks by picking the chain containing the most proof of 
work with random tie-breaking and
\item microblock forks using poison transactions~\cite{bitcoinNg}.
\end{enumerate*}

\paragraph{Censorship.} The ability of miners to block submission or retrieval 
of transactions. A key block miner becomes eligible to update the 
blockchain for a discrete epoch. However, she may 
ignore certain transactions in particular channels due to benign failures or 
malicious behavior. The extend of such censorship is limited to the miner's 
epoch, which can be adjusted by changing the key block frequency.

An adversary can leave a victim unable to retrieve transactions by controlling 
all of its connections~\cite{Heilman2015eclipse} or delaying the delivery of 
valid transactions to her~\cite{gervais2015tampering}. Countermeasures to 
mitigate such attacks apply to this work, as well.

\section{Related Work}\label{related-work}

\paragraph{Federated Chains.} Sidechains~\cite{back2014enabling} allow users 
to transfer funds across blockchains. However, this leads to fragmentation of 
the hash power. A compromised sidechain makes the main chain and the other 
sidechains vulnerable. Transfers across sidechains bloat the main chain. To 
guarantee that funds will not be pruned from the corresponding chains, such 
transfers incur high latencies. Drivechain~\cite{drivechain} attempts to 
minimize the impact of sidechains on the main chain regarding the required 
knowledge and effort to prove validity of transfers. However, this approach 
does not address inherent limitations regarding the security of sidechains.

\paragraph{Multiple Services in Bitcoin's Blockchain.} Bitcoin permits 
storage of arbitrary data on its blockchain using \texttt{OP\_RETURN} 
transactions~\cite{opreturn}. While there is no format requirement for the 
data, the size limit (currently 80 bytes) usually enforces users to store 
only a hash of their original content on the blockchain, which they externally 
validate~\cite{coloredCoins2016, omni2016}. This limitation imposes a critical 
tradeoff between data growth management and the diversity of services.

Users download and process the full history to validate the state of the 
existing blockchain protocols~\cite{nakamoto2008bitcoin, bitcoinNg, 
ethereum}. Using commodity hardware, this bootstrapping process takes many 
hours in Bitcoin~\cite{croman2016scaling}. Such protocols force users to handle 
the complexity of irrelevant services. Therefore, a monolithic history is not a 
viable option for scaling blockchains with multiple services.

\paragraph{Outsourcing the Security.} Services with distinct blockchains 
attempt to improve their security with merged mining~\cite{mergedmining} 
and anchoring.

In merged mining, a blockchain with insufficient mining power accepts proof 
of work submissions from a designated parent chain. This approach raises 
three issues.
First, if a miner is already part of the parent blockchain, she can use her 
mining power to attack the merged-mined blockchain at no cost.
Second, a merged-mined blockchain becomes dependent on its parent chain, 
making it fragile with respect to changes in the parent's security.
Finally, it is non-trivial to maintain the miner coordination across 
blockchains. Ali et al.~\cite{ali2016blockstack} show that even the largest 
merged-mined cryptocurrency, NameCoin~\cite{namecoin}, suffers from a 
single merged mining pool whose mining power exceeds the 51\% threshold.

Anchoring relies on periodically submitting the cumulative hash of all data, 
such as the root of a Merkle tree, to a trusted publishing medium, such as 
the blockchain of Bitcoin. Anchoring bloats the external blockchain and 
becomes dependent on its security.

\paragraph{Sharding the Same Service.} Elastico is a service-agnostic 
protocol for sharding blockchains~\cite{luu2016secure}. This approach 
assigns miners to committees for serializing transactions using a classical 
Byzantine consensus protocol. As in anchoring, a final committee creates a 
cumulative digest based on all shards and broadcasts it to the network.
However, to prevent double spends, Elastico requires splitting up the 
payment functionality into as many sub-services as the number of shards, 
which effectively means as many cryptocurrencies.

Treechains~\cite{treechains} is a sharding idea based on restructuring a 
blockchain into a tree of blocks, where each output has a dedicated branch 
to spend. However, this proposal is at an early stage with no prototype or a detailed technical analysis.

\section{Conclusion}\label{conclusion}

\Sos\ provides a means for improving the scalability and extensibility of 
blockchains with multiple services. \coin, the instantiation of this technique, 
reduces the resource requirements and the bootstrapping time to participate in 
the system. It provides trustless validation while preserving the same network 
and trust model as Bitcoin. Finally, it avoids fragmentation of the mining 
power that secures the blockchain.


{\tiny
\bibliographystyle{abbrv}
\bibliography{efelit}
}
\end{document}